\documentclass[aps,prl,reprint,groupedaddress]{revtex4-2}
\usepackage{amsmath}
\usepackage{overpic}
\newcommand{\bb}{\nu}
\newcommand{\la}{\lambda}
\newcommand{\al}{\alpha}

\newcommand{\ra}{\rightarrow}
\newcommand{\ee}{\mathrm{e}}
\newcommand{\ii}{\mathrm{i}}
\newcommand{\pdhfrac}[2]{\mathchoice{\frac{#1}{#2}}{#1/#2}{#1/#2}{#1/#2}}
\newcommand{\fdd}[2]{\pdhfrac{\mathrm{d}#1}{\mathrm{d}#2}}
\newcommand{\sdd}[2]{\pdhfrac{\mathrm{d}^2#1}{\mathrm{d}#2^2}}
\newcommand{\pd}[2]{\pdhfrac{{\partial}#1}{{\partial}#2}}
\newcommand{\spd}[2]{\pdhfrac{\partial^2#1}{{\partial}#2^2}}

\newcommand{\re}{\mathrm{Re}}
\newcommand{\im}{\mathrm{Im}}
\newcommand{\beq}{\begin{equation}}
\newcommand{\eeq}{\end{equation}}
\newcommand{\beqa}{\begin{eqnarray}}
\newcommand{\eeqa}{\end{eqnarray}}
\newcommand{\beqas}{\begin{eqnarray*}}
\newcommand{\eeqas}{\end{eqnarray*}}
\renewcommand{\d}[1]{\mathrm{d}#1}

\newcommand{\Ve}{V_G}
\newcommand{\Vs}{V_{\mathrm{exp}}}
\newcommand{\Us}{U_{\mathrm{exp}}}
\newcommand{\Ws}{W_{\mathrm{exp}}}
\usepackage{todonotes}

\begin{document}

\title{A Normal Form for the Onset of Collapse: \\
the Prototypical Example of the Nonlinear Schr{\"o}dinger Equation}

\author{S Jon Chapman}
\affiliation{Mathematical Institute, University of Oxford, AWB, ROQ, Woodstock Road, Oxford OX2 6GG}

\author{M. Kavousanakis}
\affiliation{School of Chemical Engineering, National Technical University of Athens, 15780, Athens, Greece}

 \author{I.G. Kevrekidis}
\affiliation{Department of Chemical and Biomolecular Engineering \& \\
Department of Applied Mathematics and Statistics, Johns Hopkins University, Baltimore, MD 21218, USA}

\author{P.G. Kevrekidis}
\affiliation{Department of Mathematics and Statistics, University of
  Massachusetts, Amherst MA 01003-4515, USA}
\affiliation{Mathematical Institute, University of Oxford, AWB, ROQ, Woodstock Road, Oxford OX2 6GG}

\date{\today}

\begin{abstract}
The study of nonlinear waves that collapse in finite time is a theme
of universal interest, e.g. within optical, atomic, plasma physics,
and nonlinear dynamics. Here we revisit the quintessential example
of the nonlinear Schr{\"o}dinger equation and systematically derive
a normal form for the emergence of blowup solutions from stationary
ones. While this is an extensively studied problem, such a normal form,
based on the
methodology of asymptotics beyond all algebraic orders,  unifies
both the dimension-dependent and power-law-dependent bifurcations
previously studied;  it yields excellent agreement with numerics
in both leading and higher-order effects;  it is applicable to both infinite
and finite domains;  and  it is valid in all (subcritical, critical and supercritical) regimes.
\end{abstract}

\maketitle

{\it Introduction.} The nonlinear Schr{\"o}dinger (NLS) model~\cite{ablowitz,ablowitz1,sulem,siambook}
has, arguably, been one of the most central nonlinear
partial differential equations (PDEs) within Mathematical Physics for the last
few decades. Its wide appeal stems from the fact that
it is a ubiquitous
envelope wave equation  arising in
a variety of diverse physical contexts.
Its applications span water waves~\cite{ir,mjarecent,dauxois},
nonlinear optical media~\cite{hasegawa,kivshar},
plasma physics~\cite{plasmas}
and more recently
the atomic physics realm
of Bose-Einstein condensates and their variants~\cite{stringari,pethick}.

The solitonic waveforms of the NLS model have been
central to all of the above investigations. A similarly prominent feature of the
NLS model
is its finite-time, self-similar blowup in higher (integer)
dimensions or for higher
powers of the associated nonlinearity.
Indeed, the latter manifestation of lack of well-posedness has been
central to both books~\cite{sulem,fibich0,boyd} and reviews~\cite{fibich,berge,pelin}
and has been the objective of continued study not only in the physical
literature, but also in the mathematical one; see,
e.g.,~\cite{pavel,pavel2,gadi} and~\cite{koch,sveta} for only some
recent examples (and also references therein). Importantly for our purposes,
these focusing aspects have become accessible to physical experiments.
On the one hand, there is the well-developed field of nonlinear optics,
where not only the well-known, two-dimensional collapsing waveform
of the Townes soliton has been observed~\cite{moll1},
but also more elaborate
themes have been touched upon including the collapse of optical vortices~\cite{vortex},
the loss of phase information of collapsing filaments~\cite{phase} or the
manipulation of the medium to avert optical collapse~\cite{psaltis}.
On the other hand, there is the flourishing area of Bose-Einstein
condensates where the Townes soliton has recently been announced~\cite{bectownes}.
Here, collapsing waveforms in higher dimensions had been experimentally identified
earlier~\cite{donley,cornish} and the ability to manipulate
the nonlinearity~\cite{haller} and the initial conditions~\cite{boris}
has continued to improve in recent times.

The emergence of collapsing
solutions out of solitonic ones is a topic that has been long
studied since the early works of~\cite{sulem1,sulem2} and summarized
in numerous reviews and books~\cite{sulem,fibich0,boyd}. Nevertheless, remarkably, a
normal form---a prototypical model equation compactly describing the relevant bifurcation, namely the
onset of
collapsing solutions out of  non-collapsing ones---does not exist,
to the best of our knowledge. Recent attempts to capture even the well-known
log-log
law of the critical case and its corrections~\cite{pavel} will confirm that.
It is known that at the critical point at which collapse emerges,
$\sigma d=2$, where $\sigma$ is the exponent of the nonlinearity and $d$
the spatial dimension of the NLS model, a symmetry enabling self-similar
rescaling of the solution towards becoming singular at a finite time (the so-called
pseudo-conformal invariance) arises.
Beyond this critical point, solitary waves become unstable
and, in a form somewhat reminiscent of the traditional pitchfork bifurcation,
 two collapsing branches of solutions emerge~\cite{yannis}. Yet, this is no ordinary pitchfork like, e.g., the one
 experimentally probed in BECs in double-well potentials~\cite{zibold}.
Here, pseudo-conformal symmetry breaks and, thus,
collapse phenomena will not follow the standard cubic pitchfork
normal form, but rather are associated with the exponentially-small,
beyond-all-algebraic-orders phenomenology of the relevant symmetry breaking.
Our aim is to go beyond the heuristic (steady state only) arguments of earlier
studies~\cite{sulem1,sulem2} and present a systematic derivation of the
associated normal form. Key features of our analysis are:

$\bullet$ We unify the case of general nonlinearity exponent and that
of arbitrary dimension, offering a result {\it broadly applicable} in the above physical settings of interest.

$\bullet$ Our analysis captures both the case of the {\it critical} log-log collapse
{\it and} the {\it supercritical} $t^{-1/2}$ {\it collapse}.

$\bullet$ Crucially, we capture not only the leading collapse order but also systematically
the {\it higher-order corrections}.

$\bullet$ We find {\it excellent agreement} with computations of the stationary solutions and
of the dynamical evolution.

\begin{figure}
  \begin{overpic}[width=0.49\textwidth]{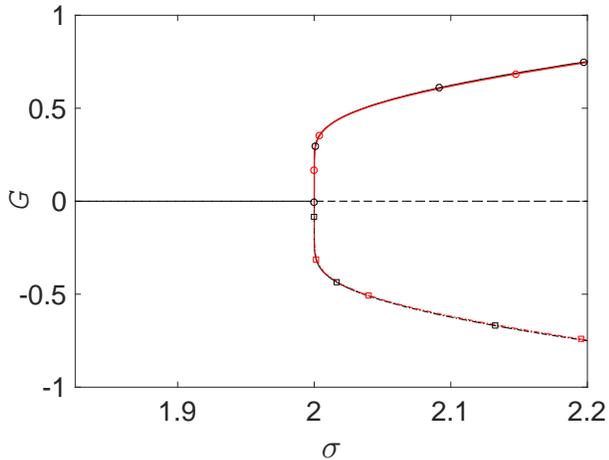}
   \end{overpic}
  \caption{Variation of the blowup rate $G$, as a function of $\sigma$ for $d=1$, domain size $K=50$. PDE results (black lines) obtained from Eq.~(\ref{mainVeqn}) are in excellent agreement with the $O(G^2)$ asymptotic solution (red lines).
 The solitonic branch ($G=0$) is stable up to $\sigma=2$ (solid line), and becomes unstable for $\sigma>2$ (dashed line).
 The stable collapsing branch ($G>0$) is depicted with solid line and open circles, and the unstable collapsing branch in the bottom ($G<0$) is illustrated with dash-dotted line and open squares.
  }
    \label{nfig1}
\end{figure}

{\it Problem Formulation \& Asymptotic Analysis.}
Upon exposing the general
formulation of the problem, we will solve it separately in the near and far
fields. The far field has a {\it turning point}, resulting in an exponentially-small reflection back towards the near field~\cite{expas}.
Matching with the near field solution yields our onset of collapse normal form,
bearing this exponentially small contribution.

We start with the NLS in dimension $d$
and nonlinearity power determined by the exponent~$\sigma$ as:
\beq
\ii \pd{\psi}{t} + \spd{\psi}{r} + \frac{(d-1)}{r}\pd{\psi}{r} + |\psi|^{2 \sigma} \psi = 0.\label{psieqn}
\eeq
We will perturb around the critical (radially symmetric) case $d \sigma=2$~\cite{sulem,fibich0}.
Introducing the well-known stretched variables~\cite{sulem,fibich0,yannis}
\[  \xi = \frac{r}{L}, \quad \tau = \int_0^t
  \frac{\d t'}{L^2(t')}, \quad \psi(r,t) = L^{-1/\sigma}
  \ee^{\ii \tau}v(\xi,\tau) \]
leads to
\beq
  \ii \pd{v}{\tau}+ \spd{v}{\xi} + \frac{(d-1)}{\xi}\pd{v}{\xi} +  |v|^{2 \sigma} v -v
  - \ii G \left(\xi \pd{v}{\xi}+ \frac{1}{\sigma} v\right)= 0,
 \label{veqn} \eeq
where the blowup rate $G=-L L_t=-L_{\tau}/L$.
In this dynamic change of variables, and in order to close the dynamics  in this ``co-exploding'' frame (upon determining $G(\tau)$),
we impose a pinning condition of the form \cite{yannis}
\[ \int_{-\infty}^\infty \re(v(\xi,\tau))T(\xi)\, \d \xi = C,\]
for some constant $C$ and some (essentially arbitrary) ``template function'' $T$, to enable us to  uniquely identify the solution $v$ and the
blowup rate $G$.
In our numerical examples we choose $T = \delta(\xi-2)$~\cite{templ}.
Finally, we write
\[ v(\xi,\tau) = V(\xi,\tau) \ee^{-\ii G(\tau) \xi^2/4}\]
to give (using $G' \equiv \fdd{G}{\tau}$)
\begin{widetext}
\beq
  \ii \pd{V}{\tau} + \frac{G' \xi^2}{4} V+ \spd{V}{\xi} + \frac{(d-1)}{\xi}\pd{V}{\xi}+ |V|^{2 \sigma}V - V
- \frac{\ii (d\sigma-2) G}{2 \sigma} V
+ \frac{G^2 \xi^2 }{4}V= 0.\label{mainVeqn}
\eeq
\end{widetext}

{\tt Near Field.} Motivated by pseudo-conformal invariance, we aim to solve \eqref{mainVeqn} in the limit $G \ra 0$ and $d \sigma \ra 2$.
We suppose   (and will verify {\em a posteriori}) that the solution evolves exponentially slowly (in $G$), and that
$\sigma$ and $d$ are exponentially close to  $\sigma_c$, $d_c$ satisfying
$d_c \sigma_c =2$~\cite{sulem,fibich0}. Thus, the second from the left and from the
right terms in Eq.~(\ref{mainVeqn})  can be neglected for now.
We look for a solution:
\beq
V =  \ee^{\ii \Phi(\tau)}\left(\Ve(\xi,\tau;G(\tau)) + \Vs(\xi,\tau)\right)
\label{extraexp}
\eeq
where $\Ve$ is the (real) regular algebraic expansion in $G$,  $\Vs$ is exponentially small in $G$, and the exponentially-slowly-varying phase $\Phi$ is determined by the pinning condition.

To obtain the near-field solution, we expand the solution in powers of $G$ as
\beq
\Ve =\sum_{n=0}^\infty G^{2n} V_n, \qquad \Phi = \sum_{n=0}^\infty G^{2n+1}\Phi_n; \label{expansion}
\eeq
this gives the leading-order equation
 \beq \spd{V_0}{\xi} + \frac{(d_c-1)}{\xi}\pd{V_0}{\xi}+ V_0^{2 \sigma_c+1} - V_0 = 0,
\label{V0eqn} \eeq
 the solution of which is the critical ground-state soliton.

The next order $V_1$ then satisfies:
\[ \spd{V_1}{\xi} + \frac{(d_c-1)}{\xi}\pd{V_1}{\xi}+ \left(\frac{4}{d_c}+1\right) V_0^{2\sigma_c} V_1 - V_1 = - \frac{\xi^2 V_0}{4},\]
with $V_1'(0) =0$, and $V_1 \ra 0 \mbox{ as } \xi \ra \infty.$

\begin{figure}
  \begin{overpic}[width=0.4\textwidth]{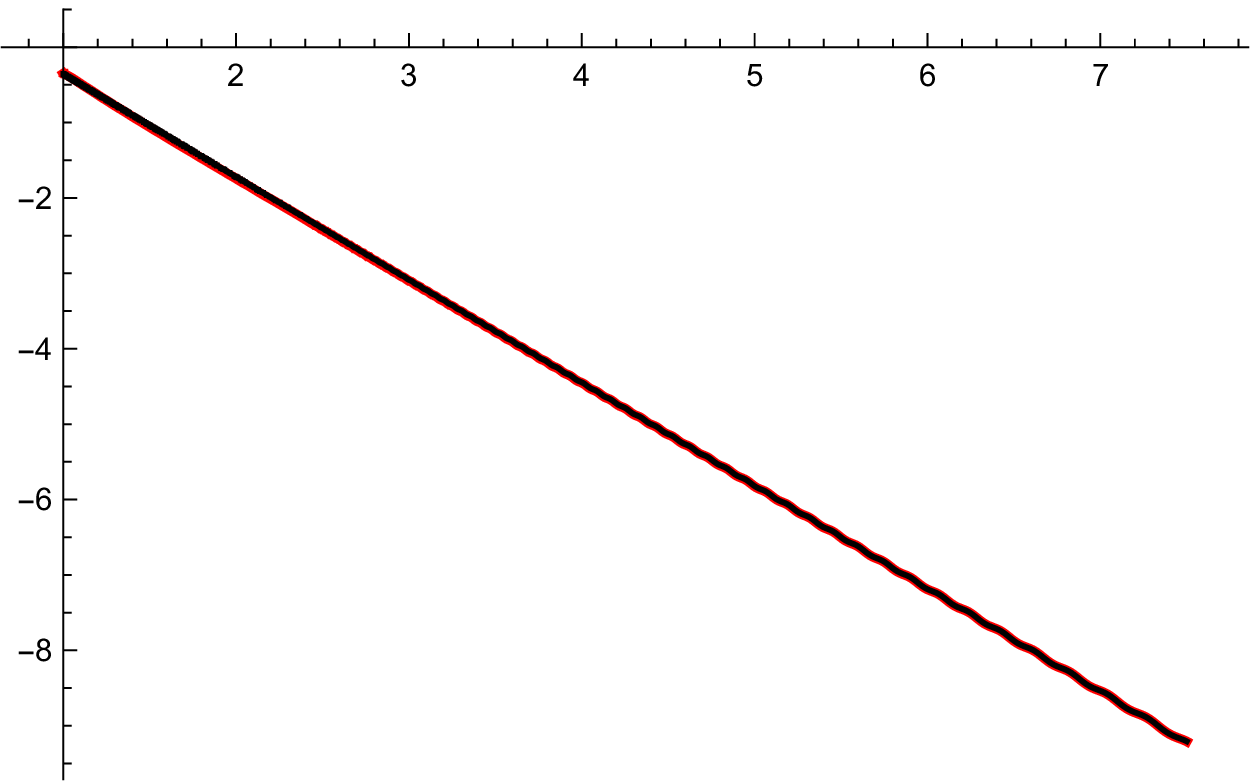}
\put(-10,10){\rotatebox{90}{$\log_{10}(G(\sigma-2))$}}
 \put(90,52){$1/G$}
  \end{overpic}
  \smallskip

   \begin{overpic}[width=0.4\textwidth]{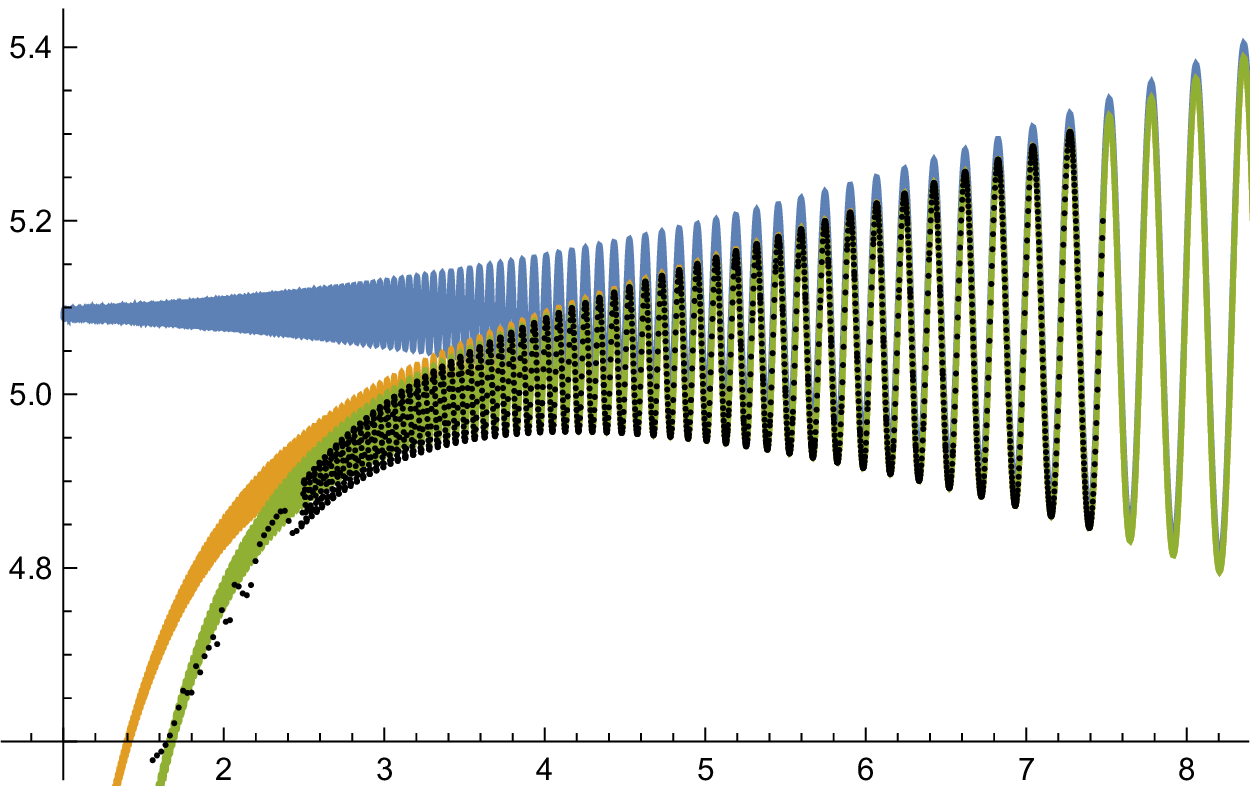}
\put(-10,10){\rotatebox{90}{$\ee^{\pi/G}G(\sigma-2)/\sigma$}}
 \put(90,-5){$1/G$}
    \end{overpic}
  \medskip
\caption{Collapsing solution branch for $d=1$, domain size $K=50$. Top panel: the leading-order asymptotic solution
    (black) is shown against a stationary numerical solution of (\ref{veqn}) (red).
    The two lines essentially coincide.
    The weak undulations are due to the sinusoidal term in (\ref{finitegamma}).
    Bottom panel: {\it exponentially scaled} illustration of the same result to show
    the accuracy of our higher-order analysis. The asymptotic solutions shown are leading order (blue),
    accurate to $O(G^2)$ (yellow), accurate to $O(G^4)$ (green);
    the  full numerical result is in black.}
    \label{nfig2}
\end{figure}

{\tt Far Field.} The above expansion (\ref{expansion}) breaks down at large distances. In the far field we rescale $\xi = \rho/G$ to give
\[
 G^2  \spd{\Ve}{\rho}+ G^2\frac{(d_c-1)}{\rho} \pd{\Ve}{\rho}+  |\Ve|^{2\sigma_c}\Ve - \Ve
 + \frac{\rho^2}{4} \Ve= 0.\]
The exponential decay of $\Ve$ renders it exponentially small in the far field,
allowing us to
neglect the nonlinear term $\Ve^{2\sigma_c+1}$.
We now look for a WKB-solution as:
\beq
\Ve \sim  G^{k}\ee^{\phi(\rho)/G} \sum_{n=0}^\infty A_n(\rho) G^{n}.\label{wkb1}
\eeq
At leading order this gives the eikonal equation:
\beq
(\phi')^2 =1-\frac{\rho^2}{4} \quad \Rightarrow\quad  \phi = -\int_0^\rho \left(1-\frac{\bar{\rho}^2}{4}\right)^{1/2}\, \d \bar{\rho}
\label{eikonal}
\eeq
(so that $\Ve$ is decreasing in $\rho$). Note the turning point at $\rho=2$
from  Eq.~(\ref{eikonal}).
The amplitude equation for $A_0$ then leads to:
\[ A_0 = \frac{a_0}{\rho^{(d_c-1)/2}(- \phi')^{1/2}} = \frac{2^{1/2}\, a_0}{\rho^{(d_c-1)/2}(4-\rho^2)^{1/4}},\]
for some constant $a_0$, which we will determine by matching with the near field.  As $\rho \ra 0$, the far field yields:
\beq
G^{k}\ee^{\phi(\rho)/G}A_0 \sim \frac{a_0G^{k}\ee^{ -\rho/G }}{\rho^{(d_c-1)/2}}  .\label{inout}
\eeq
As $\xi \ra \infty$, the near field expression is dominated by:
\beq
V_0(\xi) \sim \frac{A_{d_c} \ee^{-\xi}}{\xi^{(d_c-1)/2}} =
\frac{A_{d_c} G^{(d_c-1)/2}\ee^{-\rho/G}}{\rho^{(d_c-1)/2}} ,\label{outin}
\eeq
for some dimension-dependent constant $A_{d_c}$.
We note, in particular, the values $A_1 = 12^{1/4}$ (from the quintic NLS
exact soliton solution~\cite{sulem}), while $A_2 \approx 3.518$~\cite{pavel}.
Matching (\ref{inout}) with (\ref{outin}) gives $k = (d_c-1)/2$ and \mbox{$a_0 = A_{d_c}$}.

For $\rho>2$  only the solution of (\ref{eikonal}) in which
\[ \phi' = \ii\left(\frac{\rho^2}{4}-1\right)^{1/2}\]
has a finite Hamiltonian.
Thus for $\rho>2$,
\beq
\Ve = \al  G^{k} \ee^{\ii \phi_2(\rho)/G} \sum_{n=0}^{\infty} B_n(\rho) (\ii G)^n,\label{wkb2}
\eeq
for some constant $\alpha$, where
\[ \phi_2 = \int_2^\rho \left(\frac{\bar{\rho}^2}{4}-1\right)^{1/2}\, \d \bar{\rho}, \ \  B_0(\rho) =\frac{2^{1/2}a_0}{\rho^{(d_c-1)/2}(\rho^2-4)^{1/4}}.
\]
The fact that only one of the oscillatory exponentials is present in $\rho>2$ forces an {\it exponentially small
reflection} back towards the near field, which we will obtain by analysing the turning point region. This is a key feature of our exponential asymptotics analysis.

{\tt Turning Point.} Writing $\rho=2+G^{2/3}s$, the equation near the turning point becomes, to leading order,
\[ \sdd{\Ve}{s} + s \Ve = 0,\]
with solution
$ \Ve = \la \mbox{Ai}(-s) + \mu \mbox{Bi}(-s)$,
where Ai and Bi are Airy functions of the first and second kind, respectively.
The asymptotic expansions of Ai and Bi give:
\beqa
\Ve &\sim& \frac{\la\ee^{-2(-s)^{3/2}/3} }{2 \sqrt{\pi}(-s)^{1/4}} +
\frac{\mu \ee^{2(-s)^{3/2}/3} }{\sqrt{\pi}(-s)^{1/4}}\quad \mbox{ as }s \ra -\infty,\qquad\label{airy1}\\
\Ve & \sim & \frac{\ee^{2\ii s^{3/2}/3}}{2\sqrt{\pi}s^{1/4}} \left(\la\ee^{ - \ii \pi/4} +\mu \ee^{ \ii \pi/4}\right)\nonumber\\
&& \mbox{ }+
\frac{\ee^{-2\ii s^{3/2}/3}}{2\sqrt{\pi}s^{1/4}} \left(\la\ee^{\ii \pi/4}+\mu\ee^{-\ii \pi/4}\right)\mbox{ as }s \ra \infty.\label{airy2}
\eeqa
Matching with (\ref{wkb1}) and (\ref{wkb2}) gives $\al = \ee^{\ii \pi/4}$ and
\[ \la = \ii \mu = \frac{a_0\ii\sqrt{\pi}}{G^{1/6}}\ee^{\phi(2)/G}.
\]
Including both WKB solutions in $\rho<2$ replaces (\ref{wkb1}) with
\beq
\Ve \sim \left(\ee^{\phi(\rho)/G} + \gamma \ee^{-\phi(\rho)/G} \right)G^{k}\sum_{n=0}^\infty A_n(\rho) G^{n} ,\label{wkb1a}
\eeq
where matching with (\ref{airy1}) gives
\[ \gamma = \frac{\ii}{2}\ee^{2\phi(2)/G} = \frac{\ii}{2}\ee^{-\pi/G}.\]

\

{\tt Exponentially small correction to the near field.} As $\rho \ra 0$,
\beq
\gamma \ee^{-\phi(\rho)/G} G^{k}\sum_{n=0}^\infty A_n(\rho) G^{n} \sim \frac{  a_0 \gamma G^{(d_c-1)/2} \ee^{\rho/G}}{\rho^{(d_c-1)/2}}. \label{match2}
\eeq
This term will match with the exponentially small correction to the near field.
In the original near-field scaling, using Eq.~(\ref{extraexp})
neglecting time derivatives  and quadratic terms in $\Vs$, but keeping all the exponentially-small terms, gives
\begin{widetext}
  \beqas
\lefteqn{ \spd{\Vs}{\xi}+ \frac{(d_c-1)}{\xi}\pd{\Vs}{\xi}+ \Ve^{2 \sigma_c} \left(\sigma_c\Vs^*+(\sigma_c+1) \Vs\right)
 -\Vs
+ \frac{G^2 \xi^2 }{4}\Vs  }\qquad \qquad&& \\&& \mbox{ }
 =-\frac{(d-d_c)}{\xi} \pd{\Vs}{\xi}- \ii \pd{\Ve}{\tau}+\Phi' \Ve - \frac{G' \xi^2}{4} \Ve -2(\sigma-\sigma_c) \Ve^{2 \sigma_c+1} \log \Ve+\frac{\ii (d\sigma-2) G}{2 \sigma} \Ve,
\eeqas
where $\Phi' = \fdd{\Phi}{\tau}$. We now use $\Vs = \Us + \ii \Ws$ and separate into real and imaginary parts.
Since $\Ve$ satisfies the homogeneous version of the equation for $\Ws$, this
enables a solvability condition:
multiplying that equation by $\xi^{d_c-1} \Ve$, integrating from $0$ to $R$,
and using \eqref{V0eqn}, we obtain:
\beqa
\xi^{d_c-1} \Ve(R) \pd{\Ws}{\xi}(R) -\xi^{d_c-1} \Ws(R) \pd{\Ve}{\xi}(R)
 & = & -\int_0^R \xi^{d_c-1} \Ve \pd{\Ve}{\tau} -
\xi^{d_c-1} \frac{ (d\sigma-2) G}{2 \sigma} \Ve^2\, \d \xi
\label{solv}
\eeqa
\end{widetext}
As $R \ra \infty$ we evaluate the boundary terms by matching using (\ref{match2}), giving
\beqas
\lefteqn{\lim_{R\ra\infty}\xi^{d_c-1} \Ve(R) \pd{\Ws}{\xi}(R) -\xi^{d_c-1}  \Ws(R) \pd{\Ve}{\xi}(R)
}\quad&&\\
&\sim &
\lim_{R\ra\infty}a_0 \ee^{-R} \left( a_0 \im(\gamma) \ee^{R}\right) -  \left(a_0 \im(\gamma) \ee^{R}\right) (- a_0 \ee^{-R} )\\
& = &  2a_0^2  \im(\gamma).
\eeqas
Now
\beqas \int_0^\infty \xi^{d_c-1}\Ve^2 \, \d \xi &\sim&  \int_0^\infty \xi^{d_c-1}(V_0+G^2 V_1+ \cdots)^2 \, \d \xi\\
&\sim&
b_0 + 2 G^2 c_0+\cdots,
\eeqas
say, where
\[ b_0 =\int_0^\infty  \xi^{d_c-1} V_0^2\, \d \xi, \qquad
c_0 = \int_0^\infty  \xi^{d_c-1} V_0 V_1 \, \d \xi.\]
Thus the solvability condition (\ref{solv}) ultimately results in:
\beq 2 c_0  \fdd{G}{\tau}  =
\frac{  (d\sigma-2) b_0}{2 \sigma} -   A_{d_c}^2 \frac{\ee^{-\pi/G}}{G} .
\label{mainGeqn}
\eeq
which is the {\it normal form for the onset of collapse}.
In principle  $a_0$, $b_0$, and $c_0$ are  leading terms in  power series
expansions in $G$ above, and we can calculate the full power series. In some of our numerical examples we include the $O(G^2)$ and $O(G^4)$ corrections to these terms.
One
can discern similarities of Eq.~(\ref{mainGeqn}) with the pitchfork bifurcation normal form: the natural
bifurcation parameter is
$r=(d \sigma-2)$. Multiplying both sides by $G$, it can be
seen that for all $r<0$, $G=0$ is the only equilibrium branch of solutions.
When $r>0$, the dynamics tends towards the non-trivial (stable, collapsing) steady state solution of
Eq.~(\ref{mainGeqn}). Changing the sign of $G$ and $\tau$ and of the imaginary part $W$
(in Eq.~(\ref{mainVeqn})), we
obtain the final branch of this unusual pitchfork bifurcation diagram, a solution
that is a mirror image but is stably collapsing in negative (rather than positive) time, i.e., ``coming back from infinity''.
These are some of the intriguing by-products of unfolding the original Hamiltonian
dynamical system of Eq.~(\ref{psieqn}) into the dissipative renormalized frame of
Eq.~(\ref{mainVeqn}). Moreover, a key feature of this collapse normal form is
its exponentially small (large) nonlinear term (rather than the
usual cubic in the standard pitchfork), yielding a nearly vertical bifurcation
for $G=G(\sigma)$, as shown in Fig.~\ref{nfig1}.
Notice that our analysis is still valid for $r=0$.

\begin{figure}
\begin{overpic}[width=0.4\textwidth]{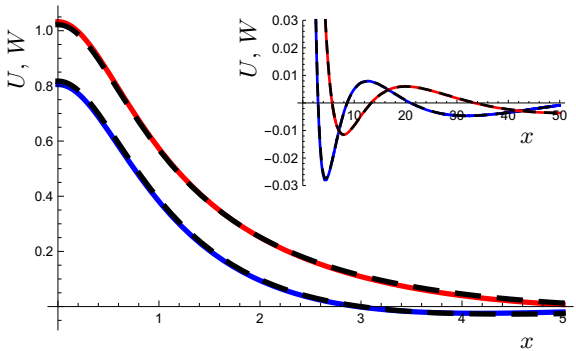}
\put(-5,45){\rotatebox{90}{$U$, $W$}}
\put(90,-3){$x$}
\put(38,47){\rotatebox{90}{$U$, $W$}}
\put(90,35){$x$}
\end{overpic}
%\vspace{6mm}
  \caption{Comparison of the numerical solution (black, dashed) with the asymptotic solution accurate to $O(G^2)$ for $K=50$, for the real ($U$, red) and the
  imaginary ($W$, blue) parts of the solution. The main plot shows the near field and the inset shows the far field. }
  \label{nfig3}
\end{figure}

{\tt Finite Domain.} Usually, when numerically simulating (\ref{psieqn}) or (\ref{mainVeqn}) the domain is truncated to some large but finite domain $[0,K]$.
In that case both oscillatory WKB solutions are present in $\rho>2$, and the ratio of their amplitudes is determined by the position of the boundary and the nature of the boundary condition. A similar analysis can be performed, and the result is a more complicated expression for the coefficient $\gamma$, the prefactor of the
reflection term at the turning point.
For example, imposing a Neumann condition on $v$ at $\xi=K$ results in
\beq
\im(\gamma)  =   \frac{(1-\bb_0^2)\ee^{- \pi/G}}
   {2(1-2\bb_0 \sin(2 \phi_2(KG)/G)) +  \bb_0^2)}, \label{finitegamma}
   \eeq
   where
   \beqas
\bb_0  =  \frac{\sqrt{(KG)^2-4} - KG}{\sqrt{(KG)^2-4}+KG}.
\eeqas
We see that as $K \ra \infty$, $\bb_0 \ra 0$ and $\im(\gamma) \ra \ee^{- \pi/G}/2$.

{\it Numerical Verification.} Equation \eqref{mainGeqn} predicts the existence of a stable branch of solutions bifurcating from $d \sigma=2$. We compare this prediction with direct  numerical simulations of \eqref{veqn} by fixing $d=1$ and varying $\sigma$ close to $\sigma_c=2$.
The relevant bifurcation diagram can be seen in Fig.~\ref{nfig1}. Here, we compare the
PDE results obtained directly from Eq.~(\ref{veqn}) with the normal form
of Eq.~(\ref{mainGeqn}) finding excellent agreement between the two.
The definitive comparison of the full NLS results with those of
our normal form is illustrated in Fig.~\ref{nfig2}. The top panel
clearly showcases the exponential nature of the relevant bifurcation
over {\it 8 orders of magnitude} of the associated ODE and PDE data in excellent
agreement between the two. Notice that the finite nature of the computation
leads to some nearly imperceptible oscillations in the top panel
of the figure, also observed but not commented in earlier works~\cite{yannis,sulem1}. The full power of our methodology is revealed
when factoring out the exponentially small leading order by rescaling through
$e^{\pi/G}$ as shown in the bottom panel of Fig.~\ref{nfig2}. In addition to the leading-order behavior we present the first- and second-order corrections, illustrating
how they progressively match in a remarkably {\it quantitative} fashion the PDE results.
To complement the quality of the match, we show in Fig.~\ref{nfig3}
how we can capture not only the rate of collapse, but also near perfectly
both the real and the imaginary parts of the profile of the associated
solution $U+\ii W$.

\begin{figure}
\begin{overpic}[width=0.49\textwidth]{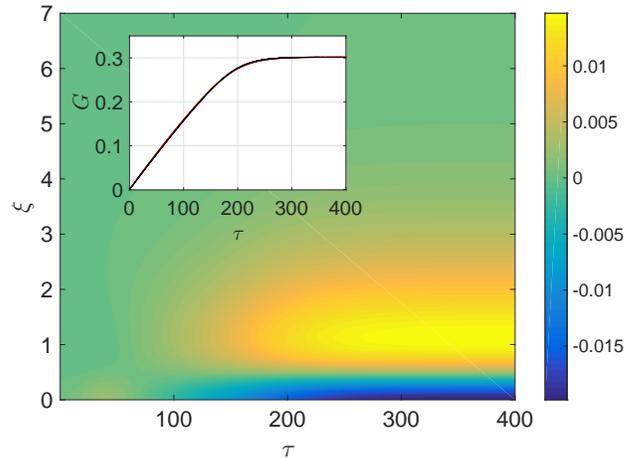}
\end{overpic}
%\vspace{6mm}
\caption{$K=50$, $\sigma=2.001$.
  Spatiotemporal evolution ($\xi - \tau$ space) of $|v|^2-|V_0|^2$.The inset shows the evolution of $G(\tau)$ for the numerical solution (red), and $O(G^4)$ asymptotic solution (black).
  The renormalized NLS reaches a steady-state solution after $\tau \approx 300$.}
\label{nfig4}
\end{figure}

Lastly, we note that our methodology not only offers a tool for capturing the
statics (i.e., the equilibrium collapse branch and its spatial profile), but
also enables an excellent capturing of the associated dynamics as shown  in Fig.~\ref{nfig4}.
Here, in addition to the spatio-temporal evolution of the field in the
$(\xi,\tau)$ variables, the evolution of the collapse rate $G(\tau)$ towards its stable asymptotic
value is observed in the inset, and compared against the numerical solution showing excellent agreement.

{\it Conclusions.} In the present work we have revisited the fundamental problem
of the collapse of a nonlinear Schr{\"o}dinger equation. We have offered a unified
perspective of the emergence of the self-similar solutions via a mathematically compact, yet
quantitatively accurate
normal form that combines the famous log-log behavior at the critical point,
the emergence of a stable self-similarly collapsing branch past that point,
the exponentially small (large) breaking of the pseudo-conformal invariance of
the critical point, the Hamiltonian nature of the original model and the
dissipative features of the renormalized dynamics. In our view this constitutes
a generic and broadly applicable (in optics, BECs and beyond)
normal form associated with the onset of collapse.
The identification of this normal form
prompts numerous exciting questions for the future, such as, e.g., the examination of
the stability of the collapsing solutions or the examination of a
potential normal form for generalized Korteweg-de Vries equations~\cite{bona}
and their traveling waves that are of broad relevance to water waves and plasmas.
This analysis may also pave the way for the study of self-similar periodic
orbits that have recently emerged in interfacial hydrodynamics~\cite{kalliadasis}.

\begin{acknowledgments}
This material is based upon work supported by the US National Science Foundation under Grants No.
PHY-1602994 and DMS-1809074 (PGK) and by the US ARO  MURI (IGK). PGK also acknowledges support from the Leverhulme Trust via
a Visiting Fellowship and thanks the Mathematical Institute of the University of Oxford for its hospitality during part of this work.
\end{acknowledgments}


\begin{thebibliography}{99}
\bibitem{templ} Naturally, there are numerous choices of the template
condition. However, the specifics of such a choice
will not affect the  normal form obtained herein.
Differentiating this template condition and substituting Eq.~(\ref{veqn}),
we obtain the algebraic equation that completes the differential-algebraic
system for $v(\xi,\tau)$ and $G$.
\end{thebibliography}

\end{document}